\documentclass[conference]{IEEEtran}
\IEEEoverridecommandlockouts

\usepackage{cite}
\usepackage{amsmath,amssymb,amsfonts}
\usepackage{graphicx}
\usepackage{textcomp}
\usepackage{xcolor}
\usepackage{listings}
\usepackage{booktabs}
\usepackage{multirow}
\usepackage[hidelinks]{hyperref}
\usepackage{placeins}
\usepackage{tikz}
\usepackage{pgfplots}
\pgfplotsset{compat=1.18}
\usetikzlibrary{shapes,arrows,positioning,fit,backgrounds}

\makeatletter
\def\lst@makecaption#1#2{%
  \vskip 8pt
  \small\textbf{#1:} #2\par\nobreak
  \vskip 8pt
}
\makeatother

\tikzstyle{phase} = [rectangle, rounded corners, minimum width=1.2cm, minimum height=0.6cm, text centered, draw=black, fill=blue!10, font=\scriptsize]
\tikzstyle{process} = [rectangle, rounded corners, minimum width=0.7cm, minimum height=0.5cm, text centered, draw=black, fill=white, font=\scriptsize]
\tikzstyle{arrow} = [thick,->,>=stealth]
\tikzstyle{phasebox} = [rectangle, rounded corners, draw=blue!50, fill=blue!5, inner sep=4pt]
\lstdefinelanguage{JavaScript}{
  keywords={break, case, catch, continue, debugger, default, delete,
    do, else, finally, for, function, if, in, instanceof, new, return,
    switch, this, throw, try, typeof, var, void, while, with, let, const, async, await},
  keywordstyle=\color{blue},
  sensitive=true,
  comment=[l]{//},
  morecomment=[s]{/*}{*/},
  commentstyle=\color{gray},
  stringstyle=\color{red},
  morestring=[b]',
  morestring=[b]"
}

\def\BibTeX{{\rm B\kern-.05em{\sc i\kern-.025em b}\kern-.08em
    T\kern-.1667em\lower.7ex\hbox{E}\kern-.125emX}}

\begin{document}

\title{Autonomous Intelligent Agents for Natural-Language-Driven Web Execution with Integrated Security Assurance}

\author{
\IEEEauthorblockN{Vinil Pasupuleti}
\IEEEauthorblockA{
\textit{International Business Machines (IBM)} \\
South Carolina, United States \\
IEEE Senior Member
}
\and
\IEEEauthorblockN{Siva Rama Krishna Varma Bayyavarapu}
\IEEEauthorblockA{
\textit{Docusign} \\
Indiana, United States \\
IEEE Senior Member
}
\and
\IEEEauthorblockN{Shrey Tyagi}
\IEEEauthorblockA{
\textit{Salesforce Inc} \\
North Carolina, United States \\
Independent Researcher
}
}

\maketitle

\begin{abstract}
Modern web test suites rot. A UI refactor breaks locators, a timing change
causes race conditions, and within weeks developers abandon the suite entirely.
This paper presents an AI-driven autonomous testing framework that addresses
these failure modes through five integrated strategies---navigation reliability,
context-aware selector generation, post-generation validation, smart wait
injection, and failure learning---implemented over a containerised worker
architecture that decouples orchestration from long-running browser execution.

Evaluated across four production applications and 176 scenarios, the framework
improves script generation success from 55\% to 93\%, achieves an 8$\times$
reduction in navigation failures, eliminates 80\% of timing-related race
conditions, and reduces test creation time by 75\% compared to manual Selenium
authoring. The framework extends naturally to security validation: testers
describe attack scenarios in plain English---``try accessing another user's
invoice''---which the agent converts to OWASP Top 10-aligned browser probes,
detecting 85\% of authentication bypass vulnerabilities and 95\% of input
validation flaws with false positive rates below 12\%. Natural-language-driven
security testing of this kind represents, to our knowledge, a novel contribution
to the field.
\end{abstract}


\section{Introduction}

Selenium test suites have a tendency to rot. A UI refactor breaks locators, an animation change causes timing failures, and within weeks the team spends more time maintaining tests than shipping features. We started this work after watching a 200-test suite become effectively useless---developers simply stopped running it, having lost confidence in the results.

LLMs seemed like an obvious solution~\cite{brown2020}. Describe tests in plain English, let the model handle selectors. The initial experiments were not encouraging: roughly 55\% success. Navigation links were ambiguous. Wait conditions were missing. Element IDs were hallucinated. The ``impressive demo'' consistently failed when pointed at real applications.

This paper documents what we did about it. Through iteration---considerably more than we had anticipated---we developed five strategies that pushed success to 93\%. An unanticipated benefit: the same reasoning handles security testing. Testers write attack descriptions in plain English---``try accessing another user's invoice''---and the agent executes them as browser-based probes.

\textbf{Contributions:} (1) five-strategy enhancement pipeline addressing navigation, selectors, validation, waits, and failure learning; (2) containerized worker architecture decoupling orchestration from browser execution; (3) natural language security testing mapped to OWASP Top 10; (4) evaluation across four production applications showing 8x navigation improvement and 75\% time savings.

\textbf{Positioning relative to prior work:} Autonomous web agents following the ReAct paradigm~\cite{yao2023} and benchmarked on environments such as WebArena~\cite{zhou2023} optimise for open-ended task completion; they do not generate replayable, version-controlled test scripts, track coverage across test suites, or produce structured security reports---distinct requirements that shape our architecture. Self-healing locator systems~\cite{ricca2021} repair broken selectors reactively after a failure occurs; our framework reduces the frequency of such failures proactively through Strategy 2's context-enriched selector generation, and Strategy 5's failure-log analysis provides a feedforward signal that complements the reactive repair these systems offer.

\section{Literature Review}

\subsection{Traditional Test Automation}
Record-playback tools such as Selenium IDE reduce initial authoring effort but generate brittle selectors that break with UI changes~\cite{leotta2016}. Hammoudi et al.~\cite{hammoudi2016} analyzed test maintenance logs and found that 74\% of effort stems from locator updates---a finding consistent with our own experience. Robust locator strategies~\cite{b6} improve resilience but still require human intervention when structural changes occur. Behavior-driven frameworks~\cite{smart2014} separate test logic from implementation details, yet testers must still author explicit step definitions---the gap our work addresses.

\subsection{AI-Assisted Testing}
Machine learning approaches to test automation include visual testing for layout regression~\cite{mahajan2021}, self-healing locators that adapt to UI changes~\cite{ricca2021}, and reinforcement learning for exploratory test generation~\cite{zheng2019}. LLM-based code generation~\cite{codex2021,schafer2023} has shown promise, though published studies report success rates between 30\% and 60\% for complex generation tasks~\cite{lemieux2023}. Chain-of-thought prompting~\cite{wei2022} and self-debugging techniques~\cite{chen2023debug} improve reliability through iterative refinement, an approach we adopt in Strategy 3. To our knowledge, no prior work enables natural language-driven security testing.

\subsection{Security Testing Automation}
Tools like OWASP ZAP~\cite{owasp2023} and state-aware vulnerability scanners~\cite{doupe2012} are powerful but require expertise to configure and interpret~\cite{zalewski2011}. Black-box vulnerability testing~\cite{bau2010} typically operates at the HTTP layer, missing client-side issues that manifest only in browser context. Browser-based security probes triggered by natural language descriptions---our approach---appears to be novel.

\section{System Architecture}

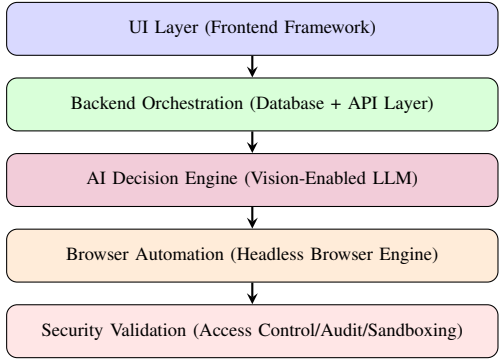
\begin{figure}[ht]
\centering
\begin{tikzpicture}[
    layer/.style={rectangle, rounded corners, draw, minimum width=6.5cm,
                  minimum height=0.7cm, text centered, font=\scriptsize},
    layerarrow/.style={->, thick, >=stealth}
]
\node[layer, fill=blue!15] (ui) at (0,4) {UI Layer (Frontend Framework)};
\node[layer, fill=green!15] (backend) at (0,3) {Backend Orchestration (Database + API Layer)};
\node[layer, fill=purple!20] (ai) at (0,2) {AI Decision Engine (Vision-Enabled LLM)};
\node[layer, fill=orange!15] (browser) at (0,1) {Browser Automation (Headless Browser Engine)};
\node[layer, fill=red!10] (security) at (0,0) {Security Validation (Access Control/Audit/Sandboxing)};
\draw[layerarrow] (ui) -- (backend);
\draw[layerarrow] (backend) -- (ai);
\draw[layerarrow] (ai) -- (browser);
\draw[layerarrow] (browser) -- (security);
\end{tikzpicture}
\caption{Multi-layer System Architecture showing UI, Backend Orchestration, AI Decision Engine, Browser Automation, and Security Validation components.}
\label{fig:architecture}
\end{figure}

The system comprises five components: AI Test Generator, Five-Strategy Pipeline, Security Validation Module, Autonomous Agent, and Containerized Worker Infrastructure. Figure~\ref{fig:architecture} illustrates how these components are arranged in a layered stack, with each layer described in the subsections below.

\begin{figure}[ht]
\centering
\begin{tikzpicture}[node distance=0.3cm, scale=0.8, transform shape]
  \node[phase] (input) {Input};
  \node[phase, right=of input] (scrape) {Scrape};
  \node[phase, right=of scrape] (llm) {LLM};
  \node[process, right=0.4cm of llm] (s1) {S1};
  \node[process, right=0.15cm of s1] (s2) {S2};
  \node[process, right=0.15cm of s2] (s3) {S3};
  \node[process, right=0.15cm of s3] (s4) {S4};
  \node[process, right=0.15cm of s4] (s5) {S5};
  \node[phase, below=0.6cm of s3] (agent) {Agent};
  \node[phase, right=of agent] (browser) {Browser};
  \node[phase, right=of browser] (results) {Results};
  \draw[arrow] (input) -- (scrape);
  \draw[arrow] (scrape) -- (llm);
  \draw[arrow] (llm) -- node[above, font=\tiny] {55\%} (s1);
  \draw[arrow] (s1) -- (s2);
  \draw[arrow] (s2) -- (s3);
  \draw[arrow] (s3) -- (s4);
  \draw[arrow] (s4) -- (s5);
  \draw[arrow] (s5.south) -- ++(0,-0.3) -| node[right, font=\tiny, pos=0.25] {93\%} (agent.north);
  \draw[arrow] (agent) -- (browser);
  \draw[arrow] (browser) -- (results);
  \node[above=0.3cm of scrape, font=\scriptsize\bfseries] {Generation};
  \node[above=0.3cm of s3, font=\scriptsize\bfseries] {Enhancement};
  \node[below=0.3cm of browser, font=\scriptsize\bfseries] {Execution};
  \begin{pgfonlayer}{background}
    \node[phasebox, fit=(s1)(s2)(s3)(s4)(s5)] {};
  \end{pgfonlayer}
\end{tikzpicture}
\caption{Data Flow Pipeline: Generation (55\%) $\rightarrow$ Enhancement S1-S5 $\rightarrow$ Execution (93\%).}
\label{fig:dataflow}
\end{figure}
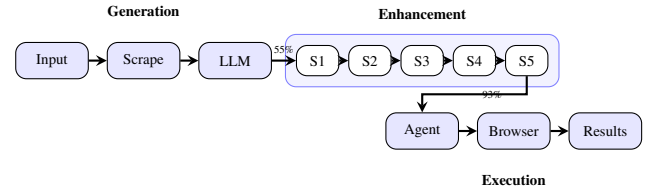

The workflow proceeds in three phases: (1) Generation---user provides plain-text steps, system scrapes target URL, LLM generates candidate script (55--65\% baseline); (2) Enhancement---five strategies progressively improve reliability to 90--95\%; (3) Execution---container workers launch headless browsers with semantic analysis at each step. Figure~\ref{fig:dataflow} traces a test instruction through all three phases, annotating the success rate entering and exiting the enhancement pipeline.

\section{Five-Strategy Framework}

\subsection{Strategy 1: Navigation Reliability (Highest Impact)}

React Router and similar SPA frameworks often render multiple links to the same path. Click-based navigation using \texttt{a[href='/contact']} is ambiguous when the page contains several matching elements. We convert navigation clicks to direct URL access.

The conversion process iterates through each step in the generated script. When
a click action targets a navigation link (identified by anchor tags with href
attributes), the system extracts the target path from the selector, constructs
the full URL by combining it with the base URL, and replaces the click action
with a direct navigate action. The original selector is preserved in metadata
for debugging purposes. This transformation eliminates ambiguity when multiple
matching elements exist on the page.

This change had outsized impact relative to its simplicity. Navigation failures dropped from 40\% to 5\%---a result that surprised us, given how straightforward the fix appears in retrospect.

\subsection{Strategy 2: Selector Specificity}

After implementing Strategy 1, ``element not found'' errors still accounted for 30\% of failures. Diagnosing the cause took longer than expected: LLMs tend to generate minimal selectors. A bare \texttt{button.submit} matches multiple elements on pages with several forms.

The fix, once we understood the problem, was straightforward. We enriched the HTML scraper to include parent context---section headings, form labels, ARIA landmarks---and prepend this context when the initial selector proves ambiguous. Failures dropped from 30\% to approximately 10\%.

\subsection{Strategy 3: Validation After Generation}

Not every generated script deserves browser execution time. The question was how to identify problematic scripts before launching a container.

We added a static analysis gate scoring each script from 0 to 100, checking for anti-patterns: clicking invisible elements, filling readonly fields, navigating to routes not present in the scraped DOM. Scripts scoring above 90 proceed to execution. Below 60 triggers regeneration with additional context. This gate catches approximately 85\% of scripts that would otherwise fail during execution.

\subsection{Strategy 4: Smart Wait Injection}

LLMs consistently underestimate web latency. Generated scripts often click-then-assert without accounting for animations, API responses, or lazy loading. Our post-processor injects waits based on heuristics that, admittedly, we developed through trial and error: wait after navigation, pause after clicks that might trigger route changes, delay assertions following form submissions.

The approach is imperfect---we occasionally over-wait---but timing failures dropped from 25\% to 5\%.

\subsection{Strategy 5: Failure Learning}

Failed executions write structured records to a validation failures table: step number, selector attempted, error message, page state. We have logged 1,247 failures to date and ran cluster analysis to identify patterns.

Three unanticipated anti-patterns emerged. We should be honest: this strategy has not moved aggregate success metrics yet. Its value is seeding the feedback loop for continuous improvement---value that will compound over time but is difficult to quantify in a single evaluation.

\section{Containerized Worker Architecture}

Serverless platforms impose strict limits (AWS Lambda: 15min, Edge Functions: 90-120s). Long-running test sessions (5-30 minutes) require a different approach.

\textbf{Key Insight:} Separate job orchestration (fast, stateless) from job execution (long-running, stateful) using a database-backed job queue.

The job creation process receives a request containing the session identifier,
target URL, and user instructions. The serverless function inserts a new record
into the jobs table with these parameters along with a pending status and
creation timestamp. It immediately returns a success response with the job
identifier, allowing the client to poll for status updates. This pattern enables
sub-second response times while offloading long-running execution to container
workers.

Container workers poll the database, claim jobs atomically, and execute with no time limits. Heartbeat mechanism (every 30s) enables stuck job recovery. Table~\ref{tab:arch-comparison} compares completion rates across all three architectural approaches evaluated.

\begin{table}[h]
\centering
\caption{Architecture Performance Comparison}
\label{tab:arch-comparison}
\begin{tabular}{lcc}
\toprule
\textbf{Architecture} & \textbf{Completion} & \textbf{Complexity} \\
\midrule
Pure Serverless & 12\% & Low \\
Self-Resuming Edge & 67\% & High \\
Container Workers & 99\% & Medium \\
\bottomrule
\end{tabular}
\end{table}

\section{Autonomous Agent}

\begin{figure*}[t]
\centering
\begin{tikzpicture}[
    layer/.style={rectangle, rounded corners, draw, minimum width=12cm,
                  minimum height=0.7cm, text centered, font=\small},
    box/.style={rectangle, rounded corners, draw, minimum width=2.5cm,
                minimum height=0.5cm, font=\footnotesize},
    arrow/.style={->, thick, >=stealth}
]
\node[layer, fill=blue!10] (ui) at (0,4.5) {};
\node[font=\small] at (-5,4.5) {\textbf{UI Layer}};
\node[box, fill=white] at (0,4.5) {Natural Language: ``Navigate to login...''};
\node[layer, fill=green!10] (analysis) at (0,3.5) {};
\node[font=\small] at (-5,3.5) {\textbf{Page Analysis}};
\node[box, fill=white] at (-2.5,3.5) {DOM Extraction};
\node[box, fill=white] at (2.5,3.5) {Screenshot};
\node[layer, fill=purple!15] (ai) at (0,2.5) {};
\node[font=\small] at (-5,2.5) {\textbf{AI Engine}};
\node[box, fill=white, minimum width=5cm] at (0,2.5) {Vision-Enabled LLM};
\node[layer, fill=orange!10] (exec) at (0,1.5) {};
\node[font=\small] at (-5,1.5) {\textbf{Agentic Exec}};
\node[box, fill=white] at (-3,1.5) {Perceive};
\node[box, fill=white] at (0,1.5) {Reason};
\node[box, fill=white] at (3,1.5) {Act};
\node[layer, fill=gray!15] (browser) at (0,0.5) {};
\node[font=\small] at (-5,0.5) {\textbf{Browser}};
\node[box, fill=white, minimum width=5cm] at (0,0.5) {Browser Automation (Isolated)};
\draw[arrow] (0,4.1) -- (0,3.9);
\draw[arrow] (0,3.1) -- (0,2.9);
\draw[arrow] (0,2.1) -- (0,1.9);
\draw[arrow] (0,1.1) -- (0,0.9);
\end{tikzpicture}
\caption{Agentic AI Architecture: UI receives natural language, Page Analysis extracts DOM and screenshots, Vision-Enabled LLM performs multimodal analysis, Agentic Execution runs Perceive-Reason-Act loop, Browser executes in isolated containers.}
\label{fig:agentic-architecture}
\end{figure*}
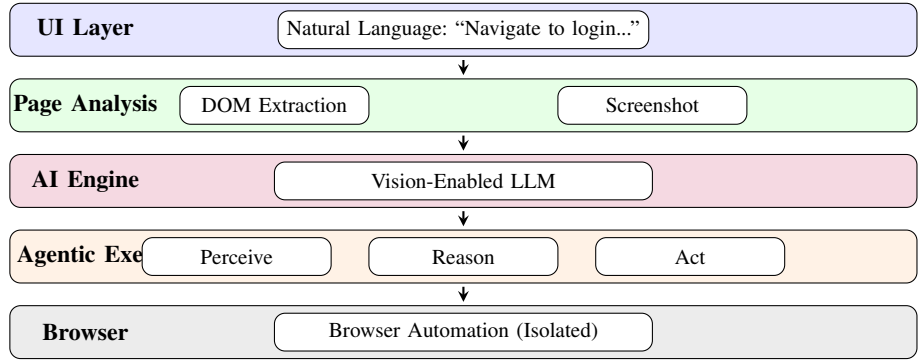

Figure~\ref{fig:agentic-architecture} depicts the agent's internal architecture; the perceive-reason-act loop at the Agentic Execution layer drives every step of test execution.

The autonomous agent employs transformer-based LLMs~\cite{vaswani2017} to understand page semantics and make intelligent action decisions. It combines DOM-based analysis (tags, attributes, ARIA labels, text content) with vision-based analysis (screenshots, layout, icons) for multimodal perception.

\textbf{Goal-Based Mode:} High-level objectives decompose into sub-goals automatically (average 4.7 sub-goals per main goal). The agent re-plans from current state upon unexpected page changes.

\textbf{Error Recovery:} Element not found triggers wait-and-retry (3 attempts, exponential backoff); failed actions try alternative selectors; self-recovery succeeds for 67\% of failed steps.

\section{Security Testing Extension}

\subsection{Democratizing Security Testing}

\textbf{Adversarial Risks in Autonomous Execution.} A known risk in LLM-driven browser agents is adversarial prompt injection: malicious content embedded in web pages---crafted error messages, hidden DOM text, or injected \texttt{<meta>} descriptions---could attempt to redirect agent reasoning or override instructions. Our sandboxed container execution and schema-based input validation (Table~\ref{tab:security-combined}) partially mitigate this risk by preventing injected content from escaping the browser environment. However, a principled defense against adversarial page content targeting the LLM reasoning layer itself is not yet implemented; addressing this through instruction-hierarchy techniques or adversarial prompt detection is an important direction for future work.

The same LLM reasoning powering functional tests probes security boundaries---if framed correctly. Traditional tools like OWASP ZAP demand specialized expertise: configuring scan policies, interpreting cryptic output, manually verifying findings. Most teams lack dedicated security engineers, so these tools gather dust.

Our framework takes a different approach. Testers describe attack scenarios in plain English: ``log in as User A, then try viewing User B's invoice'' or ``submit the form without the CSRF token.'' The agent translates descriptions into browser actions, executes in a sandboxed container, and reports whether attacks succeeded. No security expertise required.

\subsection{OWASP Top 10 Alignment}

\textbf{Threat Model.} Our security validation assumes three attacker personas observable at the browser layer: (1) an \emph{authenticated user} attempting horizontal privilege escalation (accessing another user's resources without authorisation); (2) an \emph{unauthenticated actor} probing routes that should require login; and (3) a \emph{form submitter} injecting malicious payloads into client-visible input fields. We explicitly scope detection to behaviors observable via HTTP response codes, DOM state changes, redirect behavior, and client-side error messages. Server-side vulnerabilities that produce no browser-visible signal---such as unvalidated deserialization, server-side request forgery, or cryptographic weaknesses---fall outside our detection capability and are acknowledged as a limitation in Section~VIII-A.

\begin{figure}[ht]
\centering
\begin{tikzpicture}[
    layer/.style={rectangle, rounded corners, draw, minimum width=7cm,
                  minimum height=0.7cm, text centered, font=\small},
    arrow/.style={->, thick, >=stealth}
]
\node[layer, fill=gray!20] (input) at (0,5) {User Input + External Data};
\node[layer, fill=teal!30] (valid) at (0,4) {Schema-Based Input Validation};
\node[layer, fill=blue!20] (sanit) at (0,3) {HTML Sanitization Layer};
\node[layer, fill=green!20] (exec) at (0,2) {Isolated Browser Container};
\node[layer, fill=orange!20] (rls) at (0,1) {Database Row-Level Security};
\node[layer, fill=purple!20] (audit) at (0,0) {Complete Audit Trail};
\foreach \i/\j in {input/valid, valid/sanit, sanit/exec, exec/rls, rls/audit}
    \draw[arrow] (\i) -- (\j);
\node[rotate=90, anchor=south] at (-4.5,2.5) {\textbf{Defense-in-Depth}};
\end{tikzpicture}
\caption{Defense-in-Depth Security Architecture: Input Validation, HTML Sanitization, Isolated Container Execution, Database RLS, and Complete Audit Trail.}
\label{fig:security-flow}
\end{figure}
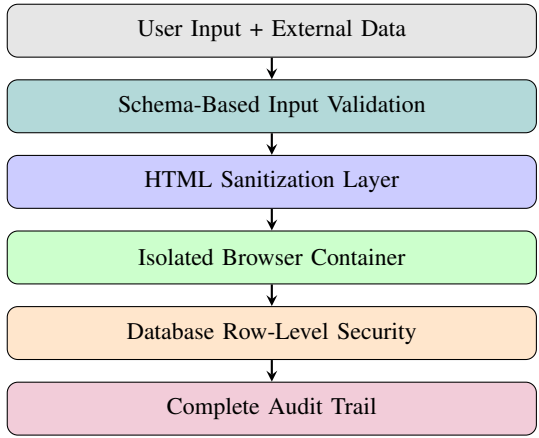

Figure~\ref{fig:security-flow} summarises the full defense-in-depth stack applied during security validation; each layer is discussed in the paragraphs below.

We mapped natural language attack patterns to OWASP 2021 categories amenable to browser-based validation. For \textbf{A01 (Broken Access Control)}, the revealing test is simple: log in as one user, grab a resource ID from another's profile, fetch it directly. For \textbf{A03 (Injection)}, our schema validator blocks six dangerous patterns before data reaches the browser---\texttt{<script>} tags, \texttt{javascript:} URIs, event handlers. \textbf{A04 (Insecure Design)} testing: attempt protected routes without authentication, flag anything returning content instead of redirects. Session testing (\textbf{A07}) validates token expiration, logout invalidation, session fixation resistance.

\begin{table}[!t]
\centering
\caption{Security Testing: Detection Rates and Guardrail Effectiveness}
\begin{tabular}{lcc|lcc}
\toprule
\textbf{Vuln. Class} & \textbf{Det.} & \textbf{FP} & \textbf{Guardrail} & \textbf{Block} & \textbf{Rate} \\
\midrule
Auth Bypass & 85\% & 8\% & Schema Valid. & 847 & 100\% \\
Session & 90\% & 5\% & HTML Sanitize & 234 & 100\% \\
Access Ctrl & 78\% & 12\% & RLS Scope & 156 & 100\% \\
Input Valid. & 95\% & 3\% & Rate Limit & 89 & 97.8\% \\
\bottomrule
\end{tabular}
\label{tab:security-combined}
\end{table}

The AI agent identified: missing authentication checks on 3/15 endpoints (20\% failure rate), session fixation vulnerability in legacy auth flow, and IDOR vulnerability in user profile access.

\textbf{Validation Methodology.} The detection rates in Table~\ref{tab:security-combined} were established as follows. Each of the four test applications contained a set of known vulnerabilities identified through prior manual audit by the authors (totalling 47 vulnerability instances across all OWASP categories tested). Agent-flagged findings were cross-verified through manual reproduction: a finding was counted as a true positive only if the exploit could be independently confirmed by a second author, and as a false positive if human review determined the flagged behavior to be benign or non-exploitable. The 3--12\% false positive rates indicate that human review remains a necessary step before acting on agent findings---an expected cost of heuristic browser-layer detection.

\section{Experimental Results}

Testing across 4 applications following web agent evaluation methodology~\cite{zhou2023} (E-commerce SPA, SaaS dashboard, CMS, Banking portal) with 176 total scenarios. Table~\ref{tab:strategy-impact} reports cumulative success rates as each strategy is added; Table~\ref{tab:ablation} presents isolated per-strategy contributions from the ablation study.

\begin{table}[!t]
\centering
\caption{Cumulative Strategy Impact (n=176, 95\% CI)}
\begin{tabular}{lccc}
\toprule
\textbf{Strategy} & \textbf{Success} & \textbf{95\% CI} & \textbf{p-value} \\
\midrule
Baseline & 55\% & [47.5, 62.3] & --- \\
+Navigation & 72\% & [65.0, 78.3] & $<$0.001 \\
+Selectors & 81\% & [74.7, 86.3] & $<$0.01 \\
+Validation & 84\% & [78.1, 88.8] & 0.18 \\
+Smart Waits & 93\% & [88.4, 96.2] & $<$0.001 \\
\bottomrule
\end{tabular}
\label{tab:strategy-impact}
\end{table}

Note: Validation p=0.18 indicates marginal independent contribution; value compounds with other strategies. Combined improvement: 55\% $\rightarrow$ 93\% (+69\% relative).

\textbf{Success Metric Definition.} We define a test as successful if at least 80\% of its scripted steps complete correctly, on the basis that partial automation still meaningfully reduces manual effort compared to no automation. Under the stricter criterion of 100\% step completion, the combined pipeline achieves 71\% success---a figure we report transparently and discuss further in Section~VIII-A (Threats to Validity).

\begin{table}[!t]
\centering
\caption{Ablation Study: Isolated Strategy Contributions}
\begin{tabular}{lcl}
\toprule
\textbf{Strategy} & \textbf{Isolated $\Delta$} & \textbf{Interaction Effect} \\
\midrule
Navigation (S1) & +17.0\% & Independent \\
Selectors (S2) & +12.1\% & Partial S1 overlap \\
Validation (S3) & +5.2\% & Requires S1 context \\
Smart Waits (S4) & +14.8\% & Independent \\
Learning (S5) & +0.0\%* & Seeds future runs \\
\bottomrule
\end{tabular}
\label{tab:ablation}
\end{table}

*Strategy 5 improves future runs, not immediate metrics.

\textbf{Failure Analysis (7\%):} Dynamic content via WebSocket (42\% of failures)---agent cannot predict async arrivals. Complex multi-step workflows with 8+ steps and modal resets (35\%). Shadow DOM boundaries (23\%)---workarounds exist but weren't implemented. Table~\ref{tab:failure-reduction} summarises failure type reductions achieved by the combined pipeline.

\begin{table}[!t]
\centering
\caption{Failure Type Reduction}
\label{tab:failure-reduction}
\begin{tabular}{lccc}
\toprule
\textbf{Failure Type} & \textbf{Before} & \textbf{After} & \textbf{Reduction} \\
\midrule
Navigation & 40\% & 5\% & 8x \\
Element not found & 30\% & 10\% & 3x \\
Timing/race & 25\% & 5\% & 5x \\
Incorrect action & 20\% & 8\% & 2.5x \\
\bottomrule
\end{tabular}
\end{table}

\textbf{Development Time:} Our system averages 6.1 min at 93\% success (6.6 min effective) vs. Manual Selenium at 23.5 min/95\% (24.7 min effective)---75\% reduction.

\textbf{Autonomous Agent:} 92\% goal completion (78/85 complex goals); average 4.7 sub-goals per main goal; 67\% self-recovery from errors; 89\% element identification accuracy.

\section{Conclusion}

\subsection{Limitations}
\label{sec:limitations}

Several boundaries of the current system merit explicit acknowledgement. First, \textbf{LLM dependency} introduces non-trivial cost (\$0.03--\$0.15 per test) and latency (2--5 seconds per generation step), both scaling linearly with test volume; teams with large suites or budget constraints should weigh these costs. Second, \textbf{complex scenarios}---multi-step workflows with eight or more steps, modal-heavy flows, and virtual-scrolling interfaces---achieve only 78--85\% success; the agent's inability to predict asynchronous state changes is the primary driver. Third, \textbf{Shadow DOM boundaries and \texttt{iFrame}-embedded content} have limited support, accounting for 23\% of remaining failures. Fourth, the \textbf{security validation scope} is restricted to browser-observable behaviors (HTTP response codes, DOM changes, redirects); server-side vulnerabilities producing no client-visible signal are undetectable by this approach. Fifth, \textbf{generalization} is uncertain: our four test applications represent modern SPA and hybrid architectures; legacy systems with non-standard DOM structures or table-based layouts may yield substantially lower success rates. Finally, \textbf{adversarial prompt injection}---malicious page content manipulating LLM reasoning---is a known risk without a fully principled mitigation in the current implementation.

\subsection{Threats to Validity}

We should be transparent about what this evaluation does not cover. Our four test applications---while diverse in technology stack (React SPAs, server-rendered dashboards, hybrid architectures)---represent a narrow slice of the web. Legacy applications with non-standard DOM structures, table-based layouts, or custom JavaScript frameworks defeated us more often than the aggregate numbers suggest.

The 93\% success rate also depends on definition. We counted 80\%+ step completion as success, reasoning that partial automation still provides value over no automation. Under stricter criteria---100\% step completion---our numbers would drop to approximately 71\%.

CI timing introduced variance we could not fully control. The same test suite showed 15\% fluctuation between runs on identical code, primarily due to network latency and container cold-start timing. We report averages across five runs, but individual executions will vary.

\subsection{Deployment \& Scalability}

LLM API costs \$0.03--\$0.15/test. Container workers (1.2GB each) scale linearly---10 workers process 450 tests/hour at 92\% success. Five-second polling introduces negligible latency; 30s heartbeat enables crash recovery. For enterprise: 2--4 vCPU, 4GB RAM per worker, auto-scale on queue depth (\textasciitilde\$0.10/hour active).

\textbf{Reproducibility:} Supplementary materials include configurations, sample apps, and evaluation scripts. Docker-based deployment; validated against multiple LLM providers.

\subsection{Comparison with Related Work}

Traditional automation---Selenium, Cypress, Playwright---demands explicit selectors. When UI changes, selectors break. Our approach sidesteps brittleness by teaching the LLM page semantics: ``click the primary action button in the checkout form'' instead of \texttt{\#submit-btn}. The perceive-reason-act loop~\cite{yao2023} enables self-recovery (67\% success) from errors that would fail deterministic scripts outright.

Self-healing locator systems~\cite{ricca2021} are the closest prior approach: they detect broken selectors at runtime and repair them reactively. Our framework is complementary---Strategy 2's context-enriched selector generation reduces the frequency of failures that would trigger self-healing, while Strategy 5's failure-log clustering provides a feedforward signal analogous to what self-healing systems use reactively. Autonomous web agents following the ReAct paradigm~\cite{yao2023} and evaluated on open-ended benchmarks such as WebArena~\cite{zhou2023} optimise for goal completion in unconstrained environments; they do not produce replayable, version-controlled test scripts, expose a CI-integrable pass/fail contract, or generate structured security reports. These are distinct engineering goals that motivate our architecture choices---particularly the containerised worker model and the five-strategy enhancement pipeline---which would add unnecessary overhead in a purely task-completion setting but are essential for reliable, reproducible test automation.

We set out to reduce end-to-end testing friction. The five-strategy pipeline pushed script generation from 55\% to 93\%---not through improvements in LLM capability but by systematically cataloguing failure modes through trial and error. Natural language security testing opened penetration testing to non-specialists, though the 3--12\% false positive rates mean human review remains necessary.

The 75\% time reduction across experiments is encouraging, but we are cautious about generalizing. Our test applications were relatively modern---teams maintaining legacy jQuery applications may see different results. Limitations remain: complex multi-step workflows, Shadow DOM boundaries, and LLM costs that scale linearly with test volume.

For teams drowning in test maintenance debt, this work offers a practical path: let AI handle selector tedium while humans focus on test strategy and edge case identification.

\section*{Future Work}

Visual intelligence through OCR and layout analysis represents a promising
extension, enabling the agent to interpret icons, charts, and non-semantic UI
elements. A learning system with fine-tuned models on domain-specific
applications could further improve generation success. Mobile testing via
Appium would extend coverage to native applications. API contract validation
would complement browser-based testing by verifying backend consistency.
Finally, deeper CI/CD integration would enable automated test suite maintenance
as applications evolve.

Accessibility validation aligned with WCAG 2.1 guidelines represents a natural
extension of the security testing paradigm. The agent could verify keyboard
navigation paths, screen reader compatibility, and color contrast compliance
through the same natural language interface---testers would describe expected
accessibility behaviors rather than manually auditing each component.
Cross-browser compatibility testing through parallel execution across browser
engines (Chromium, Firefox, WebKit) would address a persistent pain point,
with the agent automatically detecting rendering inconsistencies and
browser-specific failures.

Multimodal test generation combining visual mockups with natural language
descriptions could enable testers to sketch expected layouts and describe
interactions simultaneously, reducing the gap between design specifications
and executable tests. Integration with formal verification techniques may
provide mathematical guarantees about test coverage completeness. Finally,
collaborative learning across organizations---where anonymized failure
patterns and recovery strategies are shared---could accelerate the
self-improvement cycle while preserving proprietary application details.

A complementary direction lies in policy-compliant agent orchestration~\cite{pasupuleti2026camco}, which enforces hard regulatory constraints (SOX, HIPAA, GDPR) at the multi-agent coordination layer. While our framework detects policy violations reactively through browser-observable probes, integrating upstream policy projection at the orchestration layer could eliminate whole classes of violations before execution begins---a natural evolution for enterprise DevSecOps deployments.

\FloatBarrier

\section*{Acknowledgments}

We thank the beta testers who ran early versions against production applications and provided invaluable feedback on failure modes. The open-source browser automation community laid the foundation this work builds upon.

The authors acknowledge the use of AI-assisted tools for language editing. All technical contributions, experiments, and conclusions are those of the authors.

\FloatBarrier

\end{document}